\documentclass[prl,preprint,superscriptaddress,onecolumn,widetext]{revtex4}

\usepackage{amssymb}
\usepackage{amsmath}
\usepackage{graphicx}
\usepackage{bm}
\usepackage{amsfonts}

\newcommand{\ket}[1]{\mbox{$|#1\rangle$}}

\newcommand{\AlOOO}{\mbox{$\textrm{Al}_{2}\textrm{O}_{3}$}}

\begin{document}
\preprint{}

\title
{Decoherence in Josephson Qubits from Dielectric Loss}

\author{John M. Martinis}
\email{martinis@physics.ucsb.edu}
\author{K.B. Cooper}
\author{R. McDermott}
\author{Matthias Steffen}
\author{Markus Ansmann}
\affiliation{Department of Physics, University of California, Santa Barbara, California 93106, USA}

\author{K.D. Osborn}
\author{K. Cicak}
\author{Seongshik Oh}
\author{D.P. Pappas}
\author{R.W. Simmonds}
\affiliation{National Institute of Standards and Technology, 325 Broadway, Boulder, CO 80305-3328, USA}

\author{Clare C. Yu}
\affiliation{Department of Physics and Astronomy, University of California, Irvine, California 92697, USA}

\begin{abstract}
Dielectric loss from two-level states is shown to be a dominant
decoherence source in superconducting quantum bits. Depending on the
qubit design, dielectric loss from insulating materials or the tunnel
junction can lead to short coherence times. We show that a variety of
microwave and qubit measurements are well modeled by loss from
resonant absorption of two-level defects. Our results demonstrate that
this loss can be significantly reduced by using better dielectrics and
fabricating junctions of small area $\lesssim 10 \; \mu
\textrm{m}^2$. With a redesigned phase qubit employing low-loss
dielectrics, the energy relaxation rate has been improved by a factor
of 20, opening up the possibility of multi-qubit gates and algorithms.
\end{abstract}

\volumeyear{year}
\volumenumber{number} \issuenumber{number}
\eid{identifier}
\date{\today}

\maketitle

Superconducting qubits are a promising candidate for the construction
of a quantum computer (\textit{1}).  Circuits work well, and
experiments have demonstrated single qubit operations with reasonably
long coherence times (\textit{2 - 6}).  A recent experiment with phase
qubits (\textit{7}) has shown that the states of two qubits may be
simultaneously measured, enabling full tomographic characterization of
more complex gates (\textit{1}). Unfortunately, further progress in
this system is hindered by short coherence times. Why is the coherence
of phase qubits notably shorter than that of charge or flux qubits?
Understanding this issue will aid progress in all superconducting
qubits, as the identification of decoherence sources is crucial for
continued improvements (\textit{8 - 10}).

We report here a new decoherence mechanism, dielectric loss from
two-level states (TLS). This loss is particularly important because of
its surprisingly large magnitude; it can dominate all other sources of
decoherence. We study the dielectric loss from bulk insulating
materials as well as the tunnel barrier. A distinction is made between
these two decoherence channels, even though their fundamental source
is the same, because the losses manifest themselves differently.

This loss mechanism has been overlooked because it arises from a new
class of decoherence: it is equivalent to dissipation from a fermionic
bath (\textit{11}), which gives qualitatively different behavior
than the more familiar bosonic dissipation appropriate for photons or
phonons.  We present here several experimental measurements and a
simple TLS model that provides a detailed description of this
important phenomenon. Finally, by understanding this loss, we have
obtained in a first-generation redesign of our phase qubit a 20-fold
increase in coherence times, comparable to those of other successful
devices.

Superconducting qubits are non-linear microwave resonators formed by
the Josephson inductance of a tunnel junction and its
self-capacitance.  Coherence is destroyed by loss and noise in these
electrical elements.  For capacitors, energy loss comes from
dissipation in the insulator (with dielectric constant $\epsilon$),
which is conventionally described by the loss tangent $\tan\delta =
\textrm{Im}\{\epsilon\}/\textrm{Re}\{\epsilon\}$.  Small loss tangents
$\delta \lesssim 10^{-5}$ are desired, with the number of coherent
oscillations in the qubit given by $Q \sim 1/\delta$.

The loss tangent has generally been ignored because materials have
been assumed to exhibit low loss at low temperatures. Indeed, we and
others find that microwave loss is negligible for the crystalline
substrates Si and \AlOOO (\textit{5,12}). However, crossover wiring
in complex superconducting devices requires an insulating spacer that
is typically made from amorphous SiO$_2$ deposited by chemical vapor
deposition (CVD). In Fig. 1 we present data showing the loss tangent
of a microwave driven LC oscillator formed by a superconducting
inductor and a 300 nm thick CVD SiO$_2$ capacitor with $C \sim 5$
pF. Two small coupling capacitors connect the input and output
ports. We measure the transmitted power as a function of frequency and
drive amplitude in order to extract the loss tangent. At low
temperature $T = 25\ \textrm{mK} \ll \hbar\omega/k$, where
$\omega/2\pi \sim 6\ \textrm{GHz}$ is the resonance frequency, we find
that the loss varies strongly with drive amplitude and thus cannot be
described as a conventional resistor (bosonic bath). The loss tangent
is low at high amplitude, but scales inversely with the resonator
voltage $\langle V^2 \rangle^{1/2}$ until it saturates at an intrinsic
loss tangent $\delta_i \approx 5 \times 10^{-3}$.  Similar to high
drive powers, high temperatures give lower loss (data not shown).

Conventional measurements at high temperature or power suggest low
dielectric loss in CVD SiO$_2$.  However, superconducting qubits
operate in the $T,V \rightarrow 0$ regime, where the intrinsic loss of
SiO$_2$ is largest. Hence, great care must be taken when chosing
insulating dielectrics in any qubit design in order to prevent short
coherence times.

Dielectric loss has been previously understood to arise from resonant
absorption of microwave radiation by a bath of two-level systems
possessing an electric dipole moment (\textit{13}). The power
dependence of the loss arises from saturation of individual TLS, and
for a parallel-plate geometry is given by
\begin{align}
\label{deltalossV} \delta   &=\frac{\pi \rho (ed)^2}{3\epsilon}
\frac{\tanh(\hbar\omega/2kT)}{\sqrt{1+\omega_R^2 T_1T_2}} \ , \\
\omega_R &=(eVd/x)/\hbar \ ,
\end{align}
where $x$ is the thickness, $\rho$ is the TLS density of states each
having a fluctuating dipole moment $ed$ and relaxation times $T_1$ and
$T_2$, and $\omega_R$ is the TLS Rabi frequency.  This theory fits our
data well with parameters compatible with previous measurements of
bulk SiO$_2$ (\textit{13}).

Tunnel junctions are similarly made from amorphous dielectric
materials; are they also lossy? A key difference is that tunnel
junctions have small volume, and the assumption of a continuous
distribution of defects is incorrect. Instead, dielectric loss must be
described by a sparse bath of discrete defects. Individual defects can
be measured spectroscopically with the phase qubit (\textit{14,15}),
and in Fig. 2 we plot the peak value of the occupation probability of
the qubit $\ket{1}$ state as a function of excitation frequency and
qubit bias.  Along with the expected bias dependence, the data also
exhibit avoided two-level crossings (splittings) that arise from the
qubit state resonating with individual TLS in the tunnel barrier.
These data demonstrate the qualitative trend that small-area qubits
show fewer splittings than do large-area qubits, although larger
splittings are observed in the smaller junctions. The presence of
these spurious resonances can be quantified by measuring the amplitude
$S/h$ of each splitting, and then calculating the histogram of
amplitudes. For clarity, this distribution is better analyzed through
an integral of the number of splittings starting from the minimum
experimental resolution of 0.01 GHz.  The averages of the
corresponding integrals for seven large-area qubits and four
small-area qubits are shown in Fig. 2b. When plotted versus $\log(S)$,
we find the data fall on a line with an abrupt cutoff at
$S_\textrm{max}$, beyond which no further splittings are found.
Furthermore, the slope of the line increases with qubit junction area
$A$, and $S_\textrm{max}$ decreases with increasing $A$.

Although the splittings were initially understood as arising from
TLS fluctuations of the critical current (\textit{14}), we now
believe that charge fluctuations better describe the data
(\textit{16}). In this model, the interaction Hamiltonian
between a TLS and the qubit is given by $H_\textrm{int} =
(eVd/x)\cos\eta$, where $\eta$ is the relative angle of the dipole
moment $ed$ with respect to the electric field $V/x$. For a single
TLS dipole with two configurations L and R, the general Hamiltonian
is $2H_\textrm{TLS}= \Delta (|L\rangle\langle L| - |R\rangle\langle
R|) +\Delta_0(|L\rangle\langle R| + |R\rangle\langle L|)$.  The
eigenstates $|g\rangle = \sin(\theta/2)|L\rangle -
\cos(\theta/2)|R\rangle$ and $|e\rangle = \cos(\theta/2)|L\rangle +
\sin(\theta/2)|R\rangle$ have a energy difference
$E=\sqrt{\Delta^2+\Delta_0^2}$, where $\tan\theta =
\Delta_0/\Delta$. For a phase qubit with capacitance $C$ and a
transition energy $E_{10}$ between states $|1\rangle$ and
$|0\rangle$, the effective interaction Hamiltonian is
\begin{align}
    \label{hint} H_\textrm{int}&=i(S/2)
        (|0\rangle|e\rangle\langle 1|\langle g| -
         |1\rangle|g\rangle\langle 0|\langle e|) \ , \\
    \label{sequ} S &= S_{\textrm{max}} \cos\eta \sin\theta \ , \\
    \label{smax} S_{\textrm{max}}  &= 2\frac{d}{x}\sqrt{\frac{e^2}{2C}E_{10}} \ ,
\end{align}
where $S$ gives the size of the splitting on resonance.

The expected distribution of the splitting sizes can be calculated
using the standard TLS tunneling model, where $\Delta$ is assumed to
have a constant distribution and $\Delta_0$ has a distribution
proportional to $1/\Delta_0$ (\textit{17}).  Changing the basis to
more physical variables, the energy $E$ and the dipole matrix
element $\sin\theta$ (see Eq.(\ref{sequ})), one finds the state
density $d^2N/dEd\sin\theta \propto 1/\sin\theta \cos\theta$. An
average over $\eta$ yields
\begin{equation}\label{dos}
    \frac{d^2N}{dE dS}=\sigma A \frac{\sqrt{1-S^2/S_\textrm{max}^2}}{S}
\end{equation}
for $S<S_\textrm{max}$ and 0 otherwise, where $\sigma$ is a materials
constant describing the defect density.

This prediction is consistent with the measured splitting
distributions of Fig. 2b, where the integrated density of splittings
$dN/dE$ increases as $\log S$ until reaching a cutoff at
$S_{\textrm{max}}$.  In Fig. 2b, the thick gray trace shows a good fit
of the theory to the 13 $\mu\textrm{m}^{2}$ data using parameters
$\sigma h = 0.5 /\mu\textrm{m}^2 \textrm{GHz}$ and $S_{\textrm{max}}/h
= 0.074 \ \textrm{GHz}$.  Although the relative slope of the 70
$\mu\textrm{m}^{2}$ data scales slower than $A$, Monte Carlo
simulations confirm this arises from large resonances overlapping
smaller ones, shadowing their presence. The arrows, which indicate the
fitted values of $S_{\textrm{max}}$, agree with the scaling
$1/\sqrt{A}$ predicted by Eq. (\ref{smax}). The measured
$S_{\textrm{max}}$, along with the qubit capacitance and resonant
frequency, yield a dipole moment with $d/x \simeq 0.06$. This agrees
with previous measurements (\textit{13, 18, 19}) and makes
physical sense since $d \simeq 0.13$ nm corresponds to a single charge
moving a distance of a single atomic bond. This numerical agreement
strongly suggests that these junction resonances arise from charge TLS
and not critical-current fluctuations.

To calculate decoherence from the resonances, we first introduce a
new quantity: the average number of resonances that couple to the
qubit. An estimate of this number comes from counting the resonances
that fall within a frequency $\pm S/2$ of the qubit frequency
\begin{align}
    \label{critnumb}
    N_c &= \int_{0}^{S_{\textrm{max}}}\frac{d^2N}{dE dS}dS
    \int_{E_{10}-S/2}^{E_{10}+S/2} dE \\
        &= (\pi/4) \sigma A S_\textrm{max} \\
        &= \sqrt{A/A_c} \ ,
\end{align}
where $A_c \simeq 90$ \ $\mu\textrm{m}^2$ for $\textrm{AlO}_\textrm{x}$.
Charge and flux qubits typically have $N_c \ll 1$, whereas for phase
qubits $N_c \sim 1$.

For large-area junctions $N_c\gg 1$, the qubit couples to many
junction resonances and the decay rate $|1\rangle \rightarrow
|0\rangle$ may be calculated using the Fermi golden rule

\begin{align}
\label{gammaform}
    \Gamma_1 &=\frac{2\pi}{\hbar} \int \frac{d^2N}{dE dS}(S/2)^2 dS \\
\label{gammasa} &=(\pi/6) \sigma A S_\textrm{max}^2 /\hbar \\
           &= \frac{\pi (\sigma/x) (ed)^2}{3 \epsilon} \frac{E_{10}}{\hbar }
\end{align}
This decay rate is equivalent to a dielectric loss tangent $\delta_i
= \hbar \Gamma_1/E_{10}$ and corresponds to the $T,V \rightarrow 0$
limit of Eq. (\ref{deltalossV}).

Using Eq. (\ref{gammasa}) and the values of $\sigma$ and
$S_\textrm{max}$ determined in Fig. 2, we calculate a decay time
$1/\Gamma_1 = 8\ \textrm{ns}$ that agrees with the measured values
$10-20$ ns obtained for qubits with junction area 186
$\mu\textrm{m}^2$.  The tunnel barrier loss tangent is large $\delta_i
\simeq 1.6\times 10^{-3}$, comparable to that of CVD SiO$_{2}$. This value
is also reasonably consistent with a previous measurement of an
AlO$_{x}$ capacitor (\textit{20}).

To understand decoherence for $N_c\sim 1$, we have performed Monte
Carlo simulations of the interaction between the qubit and a
collection of resonances randomly distributed according to Eq.
(\ref{dos}). While the exact qubit dynamics depend upon the locations
of the resonances, general trends are plotted in Fig. 3.  These
simulations clearly show that the effects of dielectric loss may be
statistically avoided by designing qubits with $N_c \lesssim 1$. This
may be done by using small-area junctions or high-quality (small
$\sigma$) dielectrics for the tunnel barrier. Note that despite the
good results that are currently obtained in small area superconducting
qubits, the conclusion that the tunnel junctions are of high quality
is incorrect. The junction dielectric is actually quite lossy, but due
to the small volume it is possible to statistically avoid the discrete
nature of the loss.

We believe the large loss tangent of CVD $\textrm{SiO}_2$ and
$\textrm{AlO}_\textrm{x}$ largely explains why only a few experiments
have obtained long coherence times. For our phase qubits, junction
loss plays a prominent role in limiting the coherence for the $186 \;
\mu\textrm{m}^2$ junction. For our $70 \; \mu\textrm{m}^2$ device,
loss from SiO$_2$ is comparable to that from the junction itself,
leading to non-exponential decay of the qubit state. For our $13 \;
\mu\textrm{m}^2$ device, loss from SiO$_2$ dominates since it
contributes $\sim 10 \:\%$ to the qubit capacitance. The most
successful experiments involving charge and flux qubits
(\textit{3, 4, 21}) have used small-area junctions
\textit{and} simple designs with no lossy dielectrics directly
connected to the qubit junction, consistent with our
observations. Given the generic need for wiring crossovers in advanced
designs of a quantum computer, understanding dielectric loss is
important for future success of all qubit technologies.

The defect density of TLS, as described by the loss tangent,
determines the magnitude of decoherence.  Is it possible to lower
$\delta_i$ by improving materials?

We suggest that OH defects are the dominant source of the TLS in our
amorphous CVD $\textrm{SiO}_2$ and $\textrm{AlO}_\textrm{x}$
dielectrics. Previous experiments have measured the intrinsic loss in
undoped and doped bulk quartz at 100 to 1000 ppm concentrations
$C_\textrm{OH}$. The loss tangent was found to scale roughly as $
\delta_i \simeq 1.5\ 10^{-5} + 0.25 \ C_\textrm{OH}$
(\textit{13,22}). The $\textrm{SiO}_2$ studied in this experiment
was deposited with plasma enhanced CVD techniques using
$\textrm{SiH}_4$ and $\textrm{O}_2$ as precursor gases. A large
concentration of OH is expected for these films, on the order of a few
atomic percent (\textit{23}). The loss tangents we measure correlate
with $C_\textrm{OH}$ determined from infrared spectroscopy and agrees
in magnitude with an extrapolation of the bulk quartz data. We also
note that a previous study measured $C_\textrm{OH}$ in amorphous
$\textrm{AlO}_\textrm{x}$ to be as high as $2-8
\;\%$ (\textit{24}); this suggests why the loss tangent of
$\textrm{AlO}_\textrm{x}$ is similar to that of CVD $\textrm{SiO}_2$.

In Fig. 1 we also show dielectric loss from CVD silicon nitride, made
from precursor gases containing no oxygen.  The intrinsic loss tangent
was measured to be about 30 times smaller than for $\textrm{SiO}_2$,
again confirming the importance of reducing the OH concentration.

With SiN$_{x}$ identified as a superior dielectric, the role of
dielectric loss in phase qubits can be tested. In Fig. 4 we present
Rabi oscillation data for two phase qubits, both with 13
$\mu\textrm{m}^{2}$ area but with different wiring designs. The top
trace corresponds to our previous design with SiO$_{2}$
(\textit{15}). The bottom trace was obtained from a qubit with
SiO$_{2}$ replaced with SiN$_{x}$ and a reduction of the total amount
of dielectric.  The coherence time of the new device is about 20 times
longer than previously attained, with Rabi oscillations still visible
after 1 $\mu$s. This success gives compelling evidence that dielectric
loss plays a major role in phase qubit decoherence and defines a clear
direction for improvements in materials.

In conclusion, we have experimentally identified dielectric loss from
two-level states as an important source of decoherence in
superconducting qubits. First, we showed that loss due to common
circuit insulators such as CVD SiO$_2$ can be significant at low
temperatures and powers. We subsequently fabricated a phase qubit
device with improved insulating material, and showed an enhancement in
coherence times by a factor of 20. Second, by varying the junction
size we have shown that loss due to the tunnel junction can be
statistically avoided by fabricating small area junctions ($< 10 \;
\mu\textrm{m}^2$). By modeling the TLS as arising from discrete dipole
defects, we believe that junction resonances can also be eliminated by
a better choice of materials, possibly via removing OH defects. Our
results also clearly point toward a need for a more in-depth
understanding of dissipation due to TLS because it \emph{increases} as
$T,V \rightarrow 0$, unlike conventional bosonic dissipation channels,
and can hence lead to unexpected results. Given all of our
observations, we are optimistic that new circuits and materials can be
further developed to significantly improve the performance of all
superconducting qubits.

This work was supported by the Advanced Research and
Development Activity (ARDA) through Army Research Office grants
W911NF-04-1-2004 and MOD717304. C.C.Y. wishes to acknowledge funding
from the Deparment of Energy (DOE) through grant DE-FG02-04ER46107 and
the Office of Naval Research (ONR) through grant
N00014-04-1-0061. Some devices were fabricated at the UCSB
Nanofabrication Facility, a part of the NSF funded NNIN network.

\begin{quote}
{\bf Bibliography}

\begin{enumerate}
\item M.A. Nielsen and I.L. Chuang, {\it Quantum Computation and Quantum Information\/} (Cambridge Univ. Press, Cambridge, 2000).

\item Yu.A. Pashkin, T. Yamamoto, O. Astafiev, Y. Nakamura, D.V. Averin, J.S. Tsai, {\it Nature} {\bf 421}, 823 (2003).

\item D. Vion, A. Aassime, A. Cottet, P. Joyez, H. Pothier, C. Urbina, D. Esteve, M.H. Devoret, {\it Science} {\bf 296}, 886 (2002).

\item I. Chiorescu, Y. Nakamura, C.J.P.M. Harmans, J.E. Mooij, {\it Science} {\bf 299}, 1896 (2003).

\item A. Wallraff, D.I. Schuster, A. Blais, L. Frunzio, R.S. Huang, J. Majer, S. Kumar, S.M. Girvin, R.J. Schoelkopf, {\it Nature} {\bf 431}, 162 (2004).

\item J.M. Martinis, S. Nam, J. Aumentado, C. Urbina, {\it Phys. Rev. Lett.} {\bf 89}, 117901 (2002).

\item R. McDermott, R.W. Simmonds, M. Steffen, K.B. Cooper, K. Cicak, K.D. Osborn, S. Oh, D.P. Pappas, J.M. Martinis, {\it Science} {\bf 307}, 1299 (2005).

\item P. Bertet, I. Chiorescu, G. Burkhard, K. Semba, C.J.P.M. Harmans, D.P. DiVincenzo, J.E. Mooij, cond-mat/0412485 [{\it Phys. Rev. Lett.}, submitted] (2005).

\item E. Collin, G. Ithier, A. Aassime, P. Joyez, D. Vion, D. Esteve, {\it Phys. Rev. Lett.}, {\bf 93}, 157005 (2005).

\item O. Astafiev, Yu.A. Pashkin, Y. Nakamura, T. Yamamoto, J.S. Tsai, {\it Phys. Rev. Lett.}, {\bf 93}, 267007 (2004).

\item A. Shnirman, G. Sch\"on, I. Martin, Y. Makhlin, {\it Phys. Rev. Lett.}, {\bf 94}, 127002 (2005)

\item P.K. Day, H.G. LeDuc, B.A. Mazin, A. Vayonakis, J. Zmuidzinas, {\it Nature}, {\bf 425} 817 (2003)

\item M. v. Schickfus, S. Hunklinger, {\it Physics Letters}, {\bf 64A}, 144 (1977).

\item R.W. Simmonds, K.M. Lang, D.A. Hite, S. Nam, D.P. Pappas, J.M. Martinis, {\it Phys. Rev. Lett.}, {\bf 93}, 077003 (2004).

\item K.B. Cooper, M. Steffen, R. McDermott, R.W. Simmonds, S. Oh, D.A. Hite, D.P. Pappas, J.M. Martinis, {\it Phys. Rev. Lett.}, {\bf 93}, 180401 (2004).

\item I. Martin, L. Bulaevskii, A. Shnirman, cond-mat/0502436 [{\it Phys. Rev. Lett.}, submitted] (2005)

\item S. Hunklinger, A.K. Raychaudhuri, {\it Prog. Low. Temp. Phys.}, {\bf 9}, 265 (1986); and {\it Amorphous Solids: Low-Temperature Properties}, edited by W.A. Phillips (Springer, Berlin, 1981)

\item B. Golding, M.v. Schickfus, S. Hunklinger, K. Dransfeld, {\it Phys. Rev. Lett.}, {\bf 43}, 1817 (1979)

\item G. Baier, M.v. Schickfus, {\it Phys. Rev. B}, {\bf 38}, 9952 (1988)

\item I. Chiorescu, P. Bertet, K. Semba, Y. Nakamura, J.P.M. Harmans, J.E. Mooij, {\it Nature}, {\bf 431}, 159 (2004).

\item A. Wallraff, D.I.Schuster, A. Blais, L. Frunzio, J. Majer, S.M. Girvin, R.J. Schoelkopf, cond-mat/0502645 [{\it Phys. Rev. Lett.}, submitted] (2005).

\item M.v. Schickfus, S. Hunklinger, {\it J. Phys. C: Solid State Physics}, {\bf9}, L439 (1976)

\item V.P. Tolstoy, I. Chernyshova, V.A. Skryshevsky, {\it Handbook of Infrared Spectroscopy of Ultrathin Films} (Wiley-VCH, 2003)

\item J. Schneider, A. Andres, B. Hj\"orvarsson, I. Petrov, K. Mac\'ak, U. Helmersson, J.E. Sundgren, {\it Appl. Phys. Lett.}, {\bf 74}, 200 (1999)

\end{enumerate}
\end{quote}

\newpage

\noindent {\bf Fig. 1.} Microwave dielectric loss for materials used in
superconducting qubit fabrication.  The loss tangent $\delta$ was
determined by measuring at $T = 25$ mK the on-resonance transmission
of microwave power through a weakly coupled aluminum superconducting
LC resonator with inductance $L \approx 120$ pH and resonant frequency
between 4.7 and 7.2 GHz. Amorphous thin-film insulators of thickness
$\approx 300$ nm were grown by chemical vapor deposition and form the
dielectric of the resonator's parallel-plate capacitor.  The
a-SiO$_2$-1 (circles) and a-SiO$_2$-2 (triangles) data refer to
amorphous SiO$_2$ deposited at T=13 $^\circ$C and T=250 $^\circ$C,
respectively, using silane and oxygen as precursers.  The a- SiN$_x$
data refer to silicon nitride deposited at T=100 $^\circ$C by reacting
silane and nitrogen. As the resonator voltage approaches zero,
$\delta$ varies inversely with $\langle V^2 \rangle^{1/2}$ until
saturating at an intrinsic value $\delta_{i}$. Silicon nitride, with
$\delta_i \simeq 1.5 \times 10^{-4}$, exhibits about 30 times smaller
loss than SiO$_2$.

\noindent {\bf Fig. 2.} Measurement of junction resonances and their size
distribution.  Left: spectroscopy of the $0 \rightarrow 1$ qubit
transition as a function of junction bias for two representative phase
qubits with junction areas 13 $\mu\textrm{m}^2$ and 70
$\mu\textrm{m}^2$. When on-resonance with a TLS, the qubit spectrum
displays a splitting with magnitude $S/h$, which is proportional to
the qubit-TLS interaction energy.  The data for the larger junction,
which show more splittings that are smaller in magnitude, have been
shifted by a constant amount for clarity. Right: Size distribution of
splittings, as determined by integrating their number from size 0.01
GHz to $S'/h$, normalizing to a $1\ \textrm{GHz}$ bandwidth, and
averaging over seven (four) spectra for the 70 $\mu\textrm{m}^2$ (13
$\mu\textrm{m}^2$) qubits. Distributions are approximately linear in
log($S$) and have a sharp cutoff at $S_{\textrm{max}}$, consistent
with a fitting to the integral of Eq.(\ref{dos}) (thick gray line).
The arrows indicate the cutoff $S_{\mathrm{max}}$ and give, from
Eq.(\ref{smax}), the value $d/x=0.06$ showing that the TLS
originate in charge motion in the dielectric.

\noindent {\bf Fig. 3.} Monte Carlo simulations of a phase qubit
coupled to a bath of TLS.  The occupation probability $P_1$ of state
$\ket{1}$ is plotted versus time after initialization to the excited
state. Parameters correspond to the qubit of Fig. 2 but with varying
$N_c$. For $N_c = 5.3 $ (solid line), the qubit is coupled to many
resonances and the probability decays at a rate close to $\Gamma_1$ as
predicted by Eq.  (\ref{gammaform}). For $N_c = 1.67$ (dotted line)
the qubit is coupled to only a few resonances. The probability decays
initially with a rate $\sim\Gamma_1$, but then settles to a value near
zero that increases as $N_c$ decreases. For $N_c = 0.45$ and the qubit
tuned near a resonance (dashed line), the probability oscillates with
a large magnitude that depends on the detuning from the dominant
resonance, as discussed in ref. (\textit{15}). For $N_c = 0.45$ but
far from a resonance (dashed-dot line), the probability remains near
one and oscillates with a small magnitude that depends on the
detuning; on average the probability is $1-N_c^2/2$. These results
show that coherence dramatically improves when the qubit is designed
with $N_c \lesssim 1$.

\noindent {\bf Fig. 4.} Rabi Oscillations for a phase qubit using
  using CVD SiO$_2$ (top trace, offset) and SiN (bottom trace) as a
  dielectric for the crossover wiring.  Microwaves at the qubit
  frequency are pulsed for a time $t_R$, and subsequent measurement of
  the qubit state shows an oscillation of the probability of the
  excited state. The decay of the Rabi oscillations is consistent with
  the measured relaxation time of $T_1=0.5$ $\mu \textrm{s}$, which is
  about 20 times better than previous experiments with a SiO$_2$
  dielectric. From Ramsey fringes, we determined a dephasing time of
  $T_2 = 150$ ns (data not shown).

\clearpage

\newpage
\begin{center}
\rotatebox{0}{\put(0,0){\huge Figure 1}}
\end{center}
%\hspace{-2.5in}
%\vspace*{-5in}
\mbox{\includegraphics*[width=7.0in]{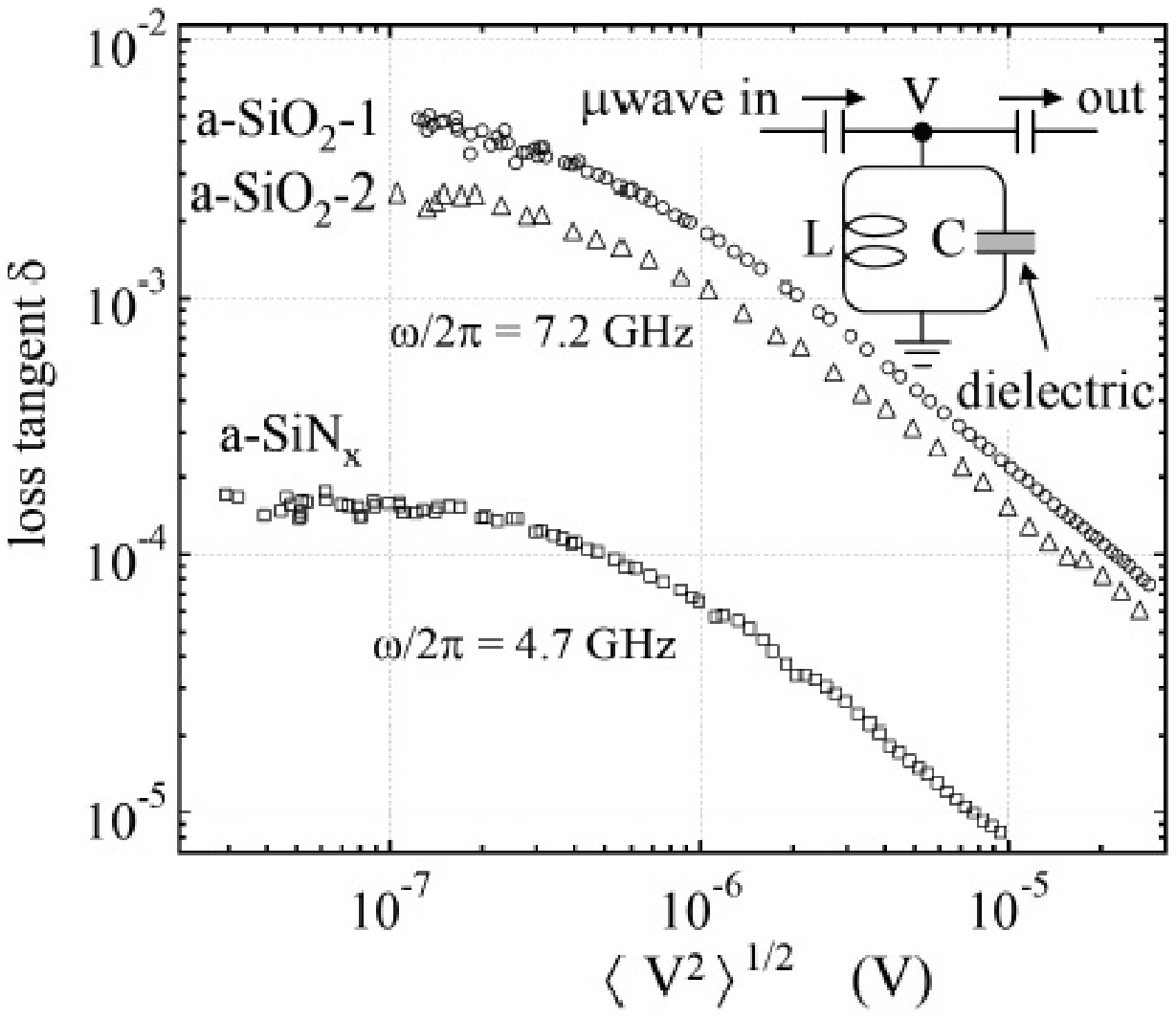}}

\newpage
\begin{center}
\rotatebox{0}{\put(0,0){\huge Figure 2}}
\end{center}
%\hspace{-2.5in}
%\vspace*{-5in}
\mbox{\includegraphics[width=7.0in]{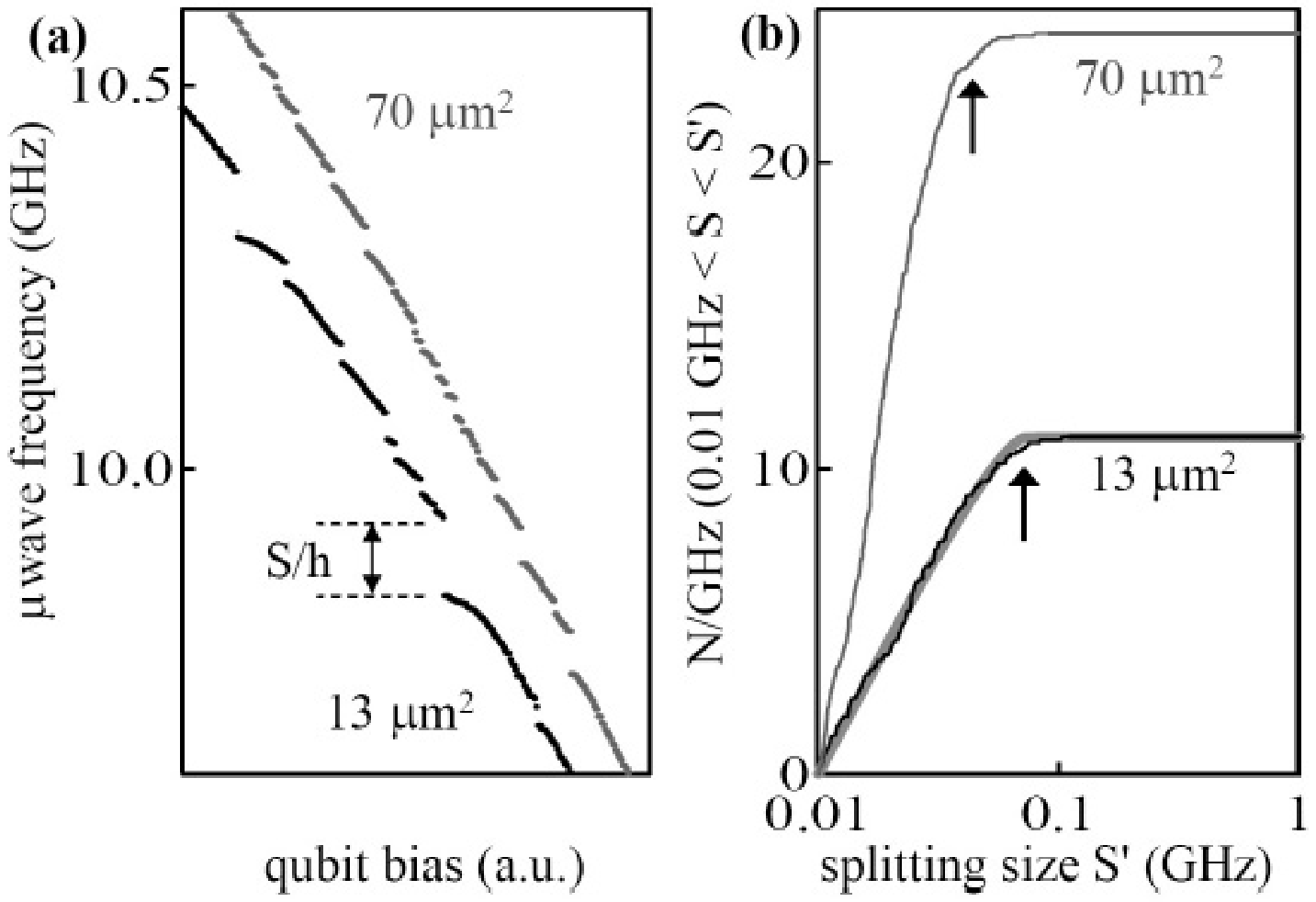}}

\newpage
\begin{center}
\rotatebox{0}{\put(0,0){\huge Figure 3}}
\end{center}
%\hspace{-4in}
%\vspace*{-7in}
\mbox{\includegraphics[width=7.0in]{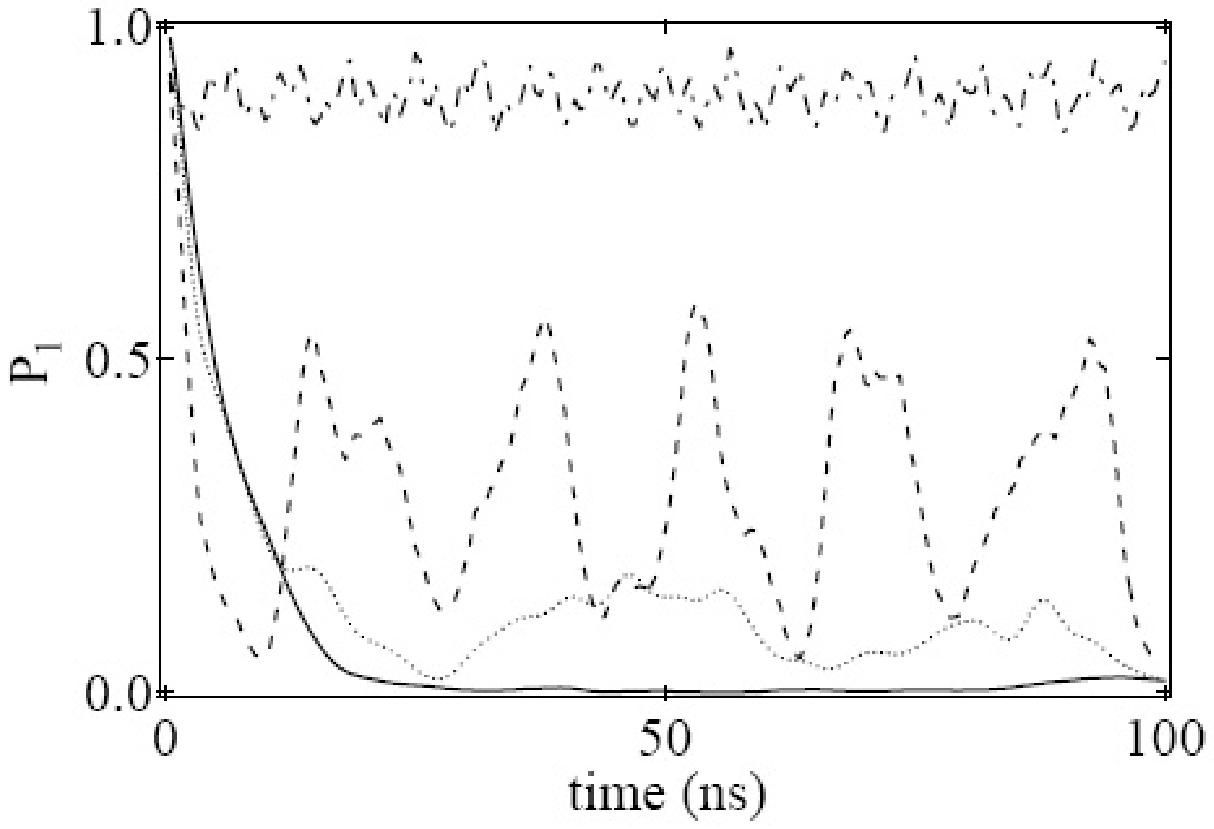}}

\newpage
\begin{center}
\rotatebox{0}{\put(0,0){\huge Figure 4}}
\end{center}
%\hspace{-4in}
%\vspace*{-7in}
\mbox{\includegraphics[width=7.0in]{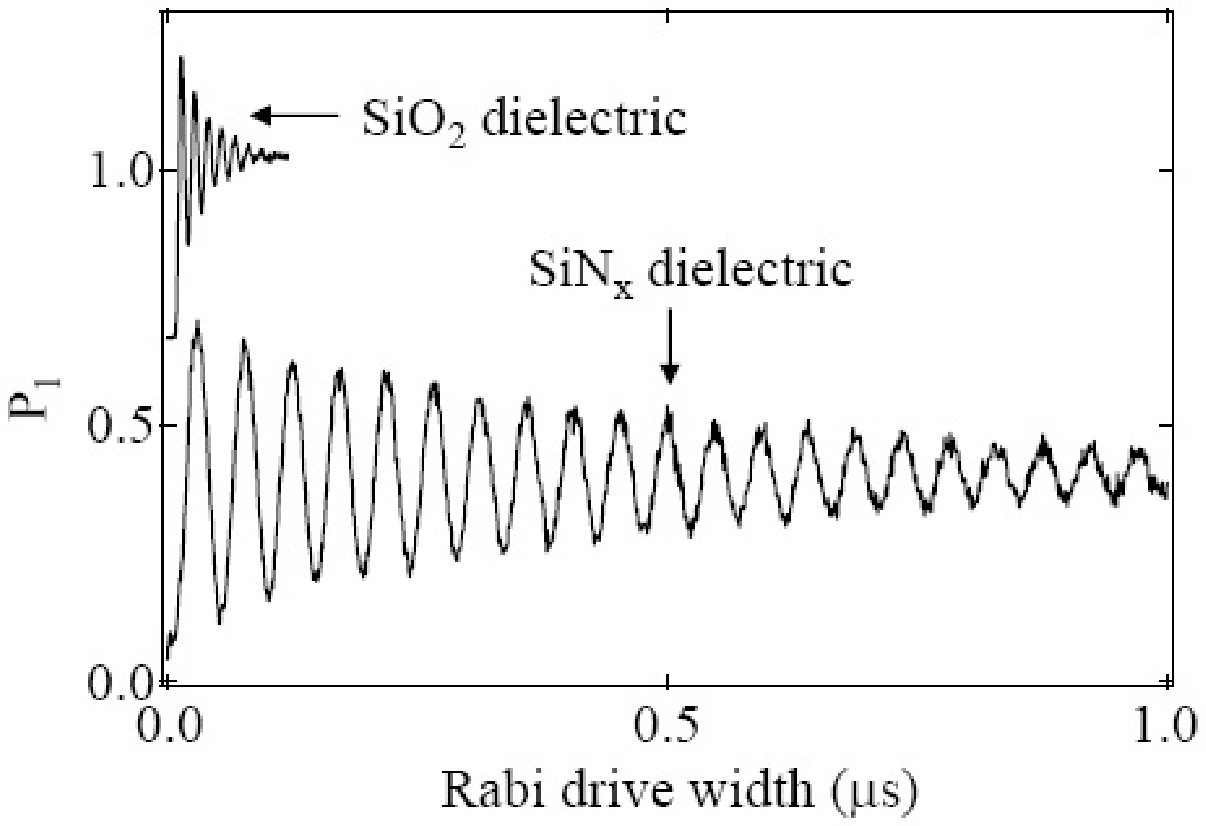}}

%\vspace*{3.5in}

\end{document}